\documentstyle[preprint,aps,psfig]{revtex}
\begin{document}

\draft
\preprint{
\vbox{
\hbox{February 1996}
\hbox{DOE/ER/40762-069}  
\hbox{U.MD. PP\#96-044}  
\hbox{ADP--95--51/T198}
}}

\title{Neutron/Proton Structure Function Ratio at Large $x$}
\author{W.Melnitchouk}
\address{Department of Physics, 
         University of Maryland, 
         College Park, MD 20742}
\author{A.W.Thomas}
\address{Department of Physics and Mathematical Physics,
         University of Adelaide,
         5005, Australia.}

\maketitle

\begin{abstract}
We re-examine the large-$x$ neutron/proton structure function 
ratio extracted from the latest deuteron data, taking into 
account the most recent developments in the treatment of Fermi motion, 
binding and nucleon off-shell effects in the deuteron.
Our findings suggest that as $x \rightarrow 1$ the 
ratio of the neutron to proton structure functions ($F_2^n/F_2^p$)
is consistent with the perturbative QCD expectation of 3/7, 
but larger than the value of 1/4 obtained in earlier analyses.
\end{abstract}

\pacs{PACS numbers: 12.38.Bx, 13.60.Hb, 25.30.Mr 	\\ \\ \\ 
      To appear in {\em Phys.Lett.B}}

% ......................................................................
\section{SU(6) Symmetry Breaking}

The precise mechanism for the breaking of the spin-flavor
SU(6) symmetry is a basic question in hadronic physics.
In a world of exact SU(6) symmetry, the wave function of a
proton, polarized say in the $+z$ direction, would be simply
\cite{CLO79}:
\begin{eqnarray}
\label{pwfn}
p\uparrow
&=& {1 \over \sqrt{2}}  u\uparrow (ud)_{S=0}\
 +\ {1 \over \sqrt{18}} u\uparrow (ud)_{S=1}\
 -\ {1 \over 3}         u\downarrow (ud)_{S=1}\ \nonumber \\
& &
 -\ {1 \over 3}         d\uparrow (uu)_{S=1}\
 -\ {\sqrt{2} \over 3}  d\downarrow (uu)_{S=1},
\end{eqnarray}
where the subscript $S$ denotes the total spin of the two-quark
component.
In this limit, apart from charge and flavor quantum numbers, 
the $u$ and $d$ quarks in the proton would be identical.  
The nucleon and $\Delta$ isobar would, for example, be degenerate 
in mass.  
In deep-inelastic scattering (DIS), exact SU(6) symmetry would be 
manifested in equivalent shapes for the valence quark 
distributions of the proton, which would be related simply by
$u_V(x) = 2 d_V(x)$ for all $x$.
For the neutron to proton structure function ratio this would imply:
\begin{eqnarray}
{ F_2^n \over F_2^p }
&=& {2 \over 3}\ \ \ \ \ \ {\rm [SU(6)\ symmetry]}.
\end{eqnarray}

In nature spin-flavor SU(6) symmetry is, of course, broken.
The nucleon and $\Delta$ masses are split by some 300 MeV.   
Furthermore, with respect to DIS, it is known that the 
$d$ quark distribution is softer 
than the $u$ quark distribution, with the neutron/proton ratio 
deviating at large $x$ from the SU(6) expectation.  
The correlation between the mass splitting in the {\bf 56} baryons 
and the large-$x$ behavior of $F_2^n/F_2^p$ was observed some 
time ago by Close \cite{CLO73} and Carlitz \cite{CAR75}.   
Based on phenomenological \cite{CLO73} and Regge \cite{CAR75}
arguments, the breaking of the symmetry in Eq.(\ref{pwfn}) was
argued to arise from a suppression of the ``diquark'' configurations
having $S=1$ relative to the $S=0$ configuration, namely   
\begin{eqnarray}
(qq)_{S=0} &\gg& (qq)_{S=1}, \ \ \ \ \ x \rightarrow 1.  
\label{S0dom}
\end{eqnarray}
Such a suppression is in fact quite natural if one observes that 
whatever mechanism leads to the observed $N-\Delta$ splitting
(e.g. color-magnetic force, instanton-induced interaction, pion
exchange), it necessarily acts to produce a mass splitting between 
the two possible spin states of the two quarks which act as spectators 
to the hard collision, $(qq)_S$, with the  
$S=1$ state heavier than the $S=0$ state by 
some 200 MeV \cite{CT}. From
Eq.(\ref{pwfn}), a dominant scalar valence diquark component 
of the proton suggests that in the $x \rightarrow 1$ limit $F_2^p$ 
is essentially given by a single quark distribution (i.e. the $u$), 
in which case:
\begin{eqnarray}
{ F_2^n \over F_2^p }
&\rightarrow& { 1 \over 4 }, \ \ \ \ \ 
{ d \over u } \rightarrow 0\ \ \ \ \
[S=0\ {\rm dominance}].
\end{eqnarray}
This expectation has, in fact, been built into most phenomenological 
fits to the parton distribution data \cite{PARAMS}.

An alternative suggestion, based on perturbative QCD, 
was originally formulated by Farrar and Jackson \cite{FJ}. 
There it was argued that the exchange of longitudinal gluons,
which are the only type permitted when the spins of the two quarks  
in $(qq)_S$ are aligned, would introduce a factor $(1-x)^{1/2}$ into 
the Compton amplitude --- in comparison with the exchange of a 
transverse gluon between quarks with spins anti-aligned.
In this approach the relevant component of the proton valence 
wave function at large $x$ is that associated with states in 
which the total ``diquark'' spin {\em projection}, $S_z$, is zero:
\begin{eqnarray}
(qq)_{S_z=0} &\gg& (qq)_{S_z=1}, \ \ \ \ \ x \rightarrow 1.
\end{eqnarray}
Consequently, scattering from a quark polarized in the opposite
direction to the proton polarization is suppressed by a factor 
$(1-x)$ relative to the helicity-aligned configuration.

A similar result is also obtained in the treatment of 
Brodsky {\em et al.} \cite{BBS} (based on counting-rules),
where the large-$x$ behavior of the 
parton distribution for a quark polarized parallel ($\Delta S_z = 1$)
or antiparallel ($\Delta S_z = 0$) to the proton helicity is given 
by:  
$q^{\uparrow\downarrow}(x) = (1~-~x)^{2n - 1 + \Delta S_z}$,
where $n$ is the minimum number of non-interacting quarks (equal 
to 2 for the valence quark distributions).
In the $x \rightarrow 1$ limit one therefore predicts:
\begin{eqnarray}
{ F_2^n \over F_2^p }
&\rightarrow& {3 \over 7}, \ \ \ \ \ 
{ d \over u } \rightarrow { 1 \over 5 }\ \ \ \ \
[S_z=0\ {\rm dominance}].
\end{eqnarray}
Note that the $d/u$ ratio {\em does not vanish} in this model.

Clearly, if one is to understand the dynamics of the nucleon's 
quark distributions at large $x$, it is imperative that the 
consequences of these models be tested against experiment.

% ......................................................................
\section{Nuclear Effects in the Deuteron}

Because of the absence of free neutron targets it is 
difficult to obtain direct data on $F_2^n$.
As a result, one usually uses a deuteron target and extracts
neutron structure information from a knowledge of the proton
structure function and the nucleon wave function in the deuteron
\cite{WEST}.
The accuracy of the extracted neutron data naturally depends on the
quality of the deuteron wave function, as well as on the extraction
procedure.

Away from the small-$x$ region ($x \agt 0.3$), the dominant 
contribution to the deuteron structure function arises from 
the impulse approximation, in which the total $\gamma^* D$ 
amplitude is factorized into $\gamma^* N$ and $N D$ amplitudes
(which one may call factorization at the amplitude level).
To order $(v/c)^2$, with $v$ the nucleon velocity, this leads 
to a convolution formula at the structure function level, 
in which the structure function of the nucleon is smeared with 
some momentum distribution, $f(y)$, of nucleons in the deuteron 
\cite{CONV,KMPW}:
\begin{eqnarray}
\label{con}
F_2^{D\ {\rm (conv)}}(x,Q^2)
&=& \int dy\ f(y)\
    F_2^N\left( {x \over y},Q^2 \right),
\label{eq:conv}
\end{eqnarray}
where $F_2^N = F_2^p + F_2^n$.
However, as explained in Refs.\cite{MST,MSTD}, 
in addition to the changes in $f(y)$ which arise when nucleon 
binding is taken into account, explicit corrections to   
Eq.(\ref{con}), which cannot be written in the form of 
a convolution, also arise when the off-mass-shell
structure of nucleons is incorporated:
\begin{eqnarray}
\label{corr}
F_2^D(x)
&=& F_2^{D\ {\rm (conv)}}(x)\
 +\ \delta^{\rm (off)} F_2^D(x).
\end{eqnarray}
Here the correction $\delta^{\rm (off)} F_2^D$ receives contributions
from the off-shell components in the deuteron wave function, and from
the off-mass-shell dependence of the bound nucleon structure function
\cite{MSTD}.
To ensure baryon number conservation, the first moment of 
$\delta^{\rm (off)} F_2^D$ is identically zero.
(For the explicit form of $f(y)$ and $\delta^{\rm (off)} F_2^D$
see Ref.\cite{MSTD}.)

The size of the non-convolution correction $\delta^{\rm (off)} F_2^D$ 
(which is relativistic in origin) was the primary aim of our previous 
study in Ref.\cite{MSTD}.  
Although model dependent, $\delta^{\rm (off)} F_2^D$ was found to  
be less than 1--2\% for most $x$ in realistic potential models of 
the $NN$ interaction \cite{BG,DREL,DNR}.
Nevertheless, in any consistent covariant treatment, it must 
be included.
Assuming in addition the conventional wisdom that 
$F_2^n/F_2^p \rightarrow 1/4$ as $x \rightarrow 1$, 
in Ref.\cite{MSTD} we also modeled the total deuteron structure
function $F_2^D$.
For the purposes of that study, the overall fit was quite 
impressive, even though on re-examination one sees that 
it was actually somewhat below the data at large $x$.

The combined effects of binding, Fermi motion and nucleon 
off-shellness on the calculated ratio $F_2^D/F_2^N$ in \cite{MSTD}
were found to be about 4--5\% at $x \sim 0.7$.  
This was qualitatively similar to the EMC effect found by 
Kaptari and Umnikov \cite{KU}, and Braun and Tokarev \cite{BT},
who used a formalism similar to ours, but did not include the 
non-convolution correction term $\delta^{\rm (off)} F_2^D$. 
On the other hand, these results differed substantially from 
earlier, on-mass-shell calculations \cite{FS78}, in which the EMC 
effect in the deuteron was only around 1\%.  
The source of these differences in the behavior of $F_2^D/F_2^N$   
may be either a difference in the neutron structure function input, 
or differences in the treatment of the nuclear effects in the 
deuteron, parametrized through the distribution function $f(y)$.
In fact, we will demonstrate that the kinematic effect of binding 
plays a critical role in the analysis, aside from any assumptions   
about the neutron structure function.

In Fig.1 we illustrate the model dependence of the ratio, $R_p$, 
of the {\em same} free proton structure function, smeared with the
function $f(y)$ calculated in the on-shell model of Ref.\cite{FS78},   
$f_{\rm on}(y)$, to that smeared with the function $f(y)$ calculated 
in Ref.\cite{MSTD} with the inclusion of binding, $f_{\rm off}(y)$:
\begin{eqnarray}
R_p(x) 
&=& { \int dy\ f_{\rm on}(y)\  F_2^p (x/y) 
\over \int dy\ f_{\rm off}(y)\ F_2^p (x/y) }.  
\end{eqnarray}
In both cases the Paris wave function \cite{DNR} has been used.  
Clearly the smearing in the on-shell model \cite{FS78} produces
a dramatically faster rise above unity for $x \agt 0.7$ than for 
the off-shell model \cite{MSTD}.  
Since the {\em same} proton structure function {\em data} 
\cite{WHIT,EMC} are used 
in both the on-shell and off-shell calculations, 
one concludes that the deviations at large $x$ arise from the 
different treatments of the kinematics.   
These differences are vital if one is interested in the large-$x$ 
behavior of the neutron structure function.

The on-shell calculation \cite{FS78} was performed in the infinite
momentum frame, where the nucleons are on their mass shells, and the
physical structure functions can be used in Eq.(\ref{con}).
One problem
with this approach is that the deuteron wave function in the infinite
momentum frame is not explicitly known.
In practice one usually makes use of the ordinary non-relativistic
$S$- and $D$-state deuteron wave functions calculated in the deuteron
rest frame \cite{DNR}. This procedure is analogous to including
only Fermi motion effects in the deuteron because 
one knows that the effect of binding in the infinite
momentum frame shows up in the presence of additional Fock components
(e.g. $NN$-meson(s) ) in the nuclear wavefunction, which 
which have not yet
been computed but which must take momentum away from the nucleons.

On the other hand, the calculation of $f(y)$ in Ref.\cite{MSTD} 
draws on the extensive experience obtained, since the discovery 
of the nuclear EMC effect, on the importance of taking into account 
binding in the treatment of the impulse approximation 
--- e.g., see Ref.\cite{ANNREV} for a recent review.
Noting the theoretical significance of the issue, and the fact that
binding has either been ignored or treated in a very approximate   
way (see below) in all published analyses of deuteron data, it is  
of some importance therefore to reanalyze the deuteron data using 
the distribution function $f(y)$ which includes the effect of binding
\cite{MSTD}, but without making any assumptions for the neutron 
structure function.

% ......................................................................
\section{Extraction of $F_2^n$}

Here we examine the consequences of analyzing the $F_2^D$ data with 
the most recent treatment of deep-inelastic scattering from the 
deuteron, in which binding and other off-shell effects
are taken into account.  
To extract the neutron structure function in a manner which is as
unambiguous as possible we shall follow the same extraction procedure 
used in previous
SLAC \cite{WHIT} and EMC \cite{EMC} data analyses, namely the smearing
(or deconvolution) method discussed by Bodek {\em et al.} \cite{BODEK}.
(For an alternative method of unfolding the neutron structure function 
see for example Ref.\cite{DECONV}.)
For completeness we briefly outline the main ingredients in this method.

Firstly, one subtracts from the deuteron data, $F_2^D$, the small, 
additive, off-shell corrections, $\delta^{\rm (off)} F_2^D$, to 
give the convolution part, $F_2^{D\ {\rm (conv)}}$.
Then one smears the proton {\em data}, $F_2^p$, with the
nucleon momentum distribution function $f(y)$ in Eq.(\ref{con})
to give $\widetilde{F}_2^p \equiv F_2^p/S_p$.
The smeared neutron structure function, $\widetilde{F}_2^n$, 
is then obtained from
\begin{eqnarray}
\label{F2nsm}
\widetilde{F}_2^n 
&=& F_2^{D\ {\rm (conv)}} - \widetilde{F}_2^p.
\end{eqnarray}
Since the smeared neutron structure function is defined as
$\widetilde{F}_2^n \equiv F_2^n/S_n$, we can invert this to
obtain the structure function of a free neutron,
\begin{eqnarray}
\label{F2n}
F_2^n 
&=& S_n \left( F_2^{D\ {\rm (conv)}} - F_2^p/S_p \right).
\end{eqnarray}

The proton smearing factor $S_p$ can be computed at each $x$ from 
the function $f(y)$, and a parametrization of the $F_2^p$ data 
(for example, the recent fit in Ref.\cite{F2PAR} to the combined
SLAC, BCDMS and NMC data).
The neutron $F_2^n$ structure function is then derived from
Eq.(\ref{F2n}) taking as a first guess $S_n = S_p$.
These values of $F_2^n$ are then smeared by the function $f(y)$,
and the results used to obtain a better estimate for $S_n$.
The new value for $S_n$ is then used in Eq.(\ref{F2n}) to obtain
an improved estimate for $F_2^n$, and the procedure repeated
until convergence is achieved.

The results of this procedure for $F_2^n/F_2^p$ are presented in
Fig.2, for both the off-shell calculation (solid) and the on-shell
model (dotted).  
The increase in the off-shell $n/p$ ratio at large $x$ can be seen 
as a direct consequence of the larger EMC effect associated with 
the ratio $R_p$ for the off-shell model shown in Fig.1.  
To illustrate the role of the non-convolution correction, 
$\delta^{\rm (off)} F_2^D$, we have also performed the analysis
setting this term to zero, and approximating $F_2^D$ by 
$F_2^{D\ {\rm (conv)}}(x)$. 
The effect of this correction (dashed curve in Fig.2) appears minimal.
One can therefore attribute most of the difference between the 
off- and on-shell results to the kinematic effect of binding in 
the calculation of $f(y)$, since both calculations involve the 
same deuteron wave functions.

The reanalyzed SLAC \cite{WHIT,GOMEZ} data points themselves are 
plotted in Fig.3, at an average value of $Q^2$ of $\approx 12$ GeV$^2$.
The very small error bars are testimony to the quality of the SLAC 
$p$ and $D$ data.   
The data represented by the open circles have been extracted with the 
on-shell deuteron model of Ref.\cite{FS78}, while the filled circles
were obtained using the off-shell model of Refs.\cite{MST,MSTD}.
Most importantly, the $F_2^n/F_2^p$ points obtained with the 
off-shell method appear to approach a value broadly consistent
with the Farrar-Jackson \cite{FJ} and Brodsky {\em et al.} \cite{BBS}
prediction of 3/7, whereas the data previously analyzed in terms
of the on-shell formalism produced a ratio that tended to the lower
value of 1/4.

The $d/u$ ratio, shown in Fig.4, is obtained by simply inverting
$F_2^n/F_2^p$ in the valence quark dominated region.
The points extracted using the off-shell formalism (solid circles) 
are again significantly above those obtained previously with the 
aid of the on-shell prescription.
In particular, they indicate that the $d/u$ ratio may actually 
approach a {\em finite} value in the $x \rightarrow 1$ limit, 
contrary to the expectation of the model of Refs.\cite{CLO73,CAR75},
in which $d/u$ tends to zero. 
Although it is {\em a priori} not clear at which scale the model 
predictions \cite{CLO73,CAR75,FJ,BBS} should be valid, for the values  
of $Q^2$ corresponding to the analyzed data the effects of 
$Q^2$ evolution are minimal.

Naturally it would be preferable to extract $F_2^n$ without 
having to deal with uncertainties in the nuclear effects.
In principle this could be achieved by using neutrino and 
antineutrino beams to measure the $u$ and $d$ distributions in 
the proton separately, and reconstructing $F_2^n$ from these.
Unfortunately, as seen in Fig.4, the neutrino data from the 
CDHS collaboration \cite{CDHS} do not extend out to very large 
$x$ ($x \alt 0.6$), and at present cannot discriminate between 
the different methods of analyzing the electron--deuteron data.

We should also note that the results of our off-shell model are
qualitatively similar \cite{GOMEZ} to those obtained using the 
nuclear density method suggested by Frankfurt and Strikman 
\cite{FS88}.
There the EMC effect in deuterium was assumed to scale with that 
in heavier nuclei according to the ratio of the respective nuclear
densities, so that the ratio $F_2^D/F_2^N$ in the trough region was
depleted by about 4\%.
While this is qualitatively a reasonable way to deal with the binding
correction, the extrapolation from heavier nuclei to the deuteron was
based on an average density approximation. Because of the special nature
of the deuteron, where the neutron and proton are on average more than
4 fm apart and there is a significant $D$-state component, such an
extrapolation cannot be considered quantitatively reliable.

On the other hand, our results contradict those of Liuti and Gross  
\cite{LG}, who have recently tried to extract $F_2^n/F_2^p$ using 
an extension of the nuclear extrapolation method of Ref.\cite{FS88} 
and the formalism of Ref.\cite{GL}, in combination with 
a non-relativistic expansion formula approximation for $F_2^D$. 
It is known, however, that the expansion formula is reliable 
only at moderate values of $x$ ($x \alt 0.7$), and indeed 
{\em overestimates} the convolution results above $x \sim 0.7$
\cite{KMPW}.
This is clear because in its derivation one has to neglect the 
lower limit on the $y$-integration in Eq.(\ref{con}), effectively
replacing $y_{min} = x$ with $y_{min} = 0$.
At very large $x$ this approximation must start to break down, 
resulting in an underestimate of $F_2^n$.
This may explain the lower values for the $n/p$ ratio obtained 
in Ref.\cite{LG}.

% ......................................................................
\section{Summary}

As explained above, earlier analyses of the large-$x$ behavior of the
neutron structure function based on deuteron data have either ignored
the effect of binding or have used a more qualitative treatment
based on an extrapolation in terms of an average density.
In view of the demonstrated importance of binding 
we have reanalyzed the latest proton and deuteron 
structure function data at large $x$, in order to obtain more 
reliable information on the structure of the neutron in the 
limit $x \rightarrow 1$.
Including all of the currently known nuclear effects in the 
deuteron, namely Fermi motion, binding, and nucleon
off-mass-shell effects, we find that the 
total EMC effect is larger than in previous 
calculations based on on-mass-shell kinematics, from which 
binding effects were essentially excluded.
This translates into an increase in the ratio 
$F_2^n/F_2^p$ at large $x$.
Our results indicate that the limiting value as $x \rightarrow 1$
is above the previously accepted result of 1/4, and broadly
consistent with the perturbative QCD expectation of 3/7.  
This also implies that the $d/u$ ratio approaches a 
{\em non-zero} value around 1/5 as $x \rightarrow 1$.

We should also point out similar consequences for the spin-dependent
neutron structure function $g_1^n$, where the Close/Carlitz
\cite{CLO73,CAR75} and Farrar/Jackson \cite{FJ,BBS} models also 
give different predictions for $g_1^n/g_1^p$ as $x \rightarrow 1$, 
namely 1/4 and 3/7, respectively.
Quite interestingly, while the ratio of polarized to unpolarized 
$u$ quark distribution is predicted to be the same in the two 
models, 
\begin{eqnarray}
{ \Delta u \over u }
&\rightarrow& 1\ \ \ \ [S=0\ or\ S_z=0\ {\rm dominance}],
\end{eqnarray}
the results for the $d$-quark distribution ratio differ even
in sign:
\begin{mathletters}
\begin{eqnarray}
{ \Delta d \over d }
&\rightarrow& - {1 \over 3}\ \ \ \ [S=0\ {\rm dominance}], 	\\
&\rightarrow& 1\ \ \ \ [S_z=0\ {\rm dominance}].
\end{eqnarray}
\end{mathletters}%

To extract information on the polarized parton densities at large $x$
that is capable of discriminating between these predictions, the same
care will need to be taken when subtracting the nuclear effects from
$g_1^D$ and $g_1^{^3He}$. 
In particular, the results of Refs.\cite{KMPW,MPT} indicate that while the 
simple prescription \cite{DPOL} of subtracting the $g_1^p$ structure
function from the $D$ data,  
modified only by the deuteron 
$D$-state probability, is surprisingly good for $x \alt 0.6$, it is 
completely inadequate for $x \agt 0.7$.

Finally, for more definitive tests of the nuclear effects in the  
deuteron, it has been suggested that one might perform a series 
of semi-inclusive experiments on deuteron targets, measuring in 
coincidence both the scattered lepton and recoiling proton or 
neutron.
Such experiments are already planned for CEBAF and HERMES 
\cite{SEMI}, and should provide critical information on the 
size and importance of relativistic and other short-distance 
nuclear effects in the deuteron.

%%%%%%%%%%%%%%%%%%%%%%%%%%%%%%%%%%%%%%%%%%%%%%%%%%%%%%%%%%%%%%%%%%%%%%%%
\acknowledgements

We would like to thank S.Rock for bringing our attention to this
problem, I.Sick for sending the deuteron data, and S.J.Brodsky 
and F.E.Close for useful discussions.  
This research was partially supported by the  
Australian Research Council and the U.S. Department of Energy 
grant \# DE-FG02-93ER-40762.

%%%%%%%%%%%%%%%%%%%%%%%%%%%%%%%%%%%%%%%%%%%%%%%%%%%%%%%%%%%%%%%%%%%%%%%%

%%%%%%%%%%%%%%%%%%%%%%%%%%%%%%%%%%%%%%%%%%%%%%%%%%%%%%%%%%%%%%%%%%%%%

\begin{figure}
\centering{\ \psfig{figure=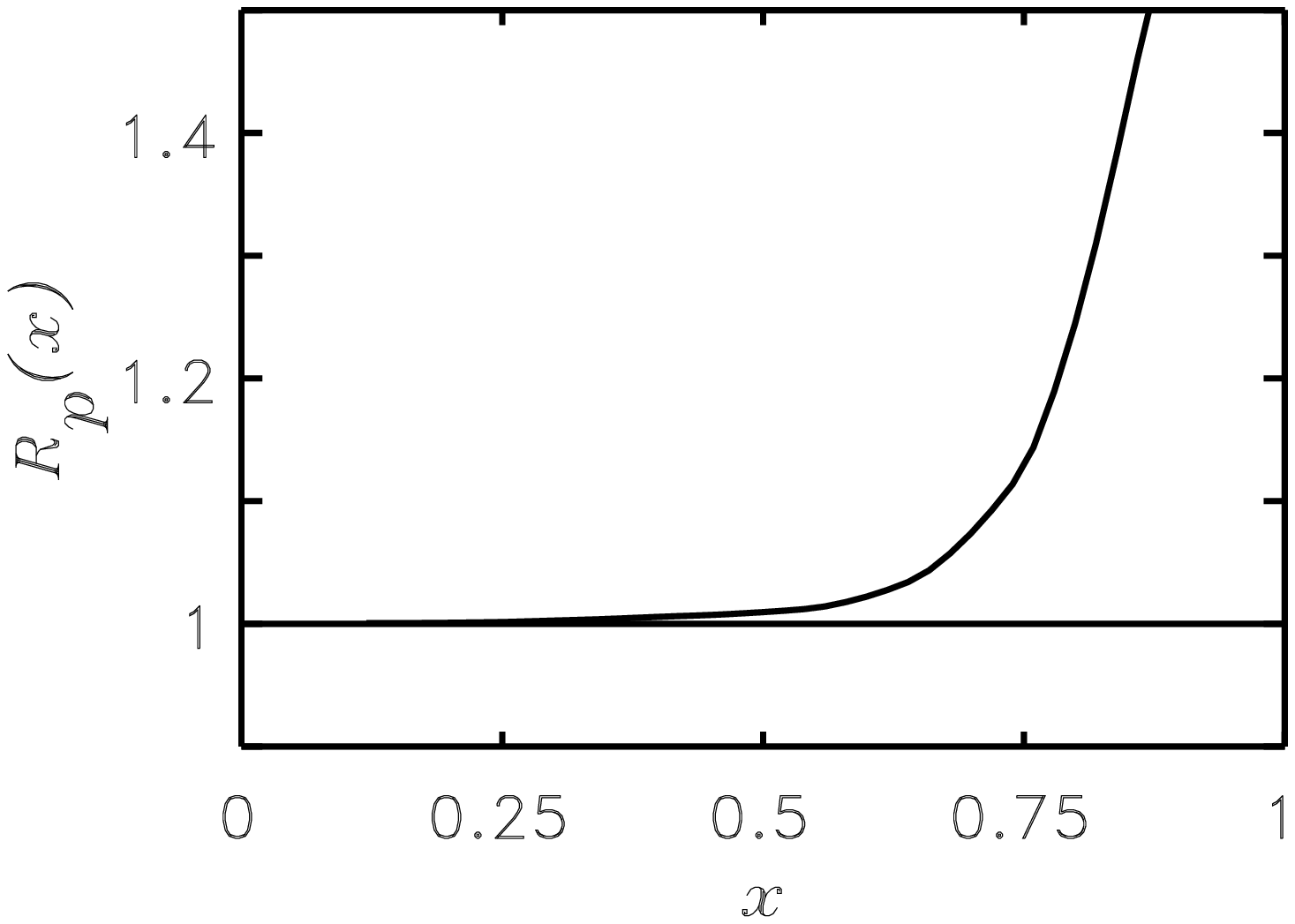,height=11cm}}
\caption{Model dependence of the ratio $R_p(x)$ of the free
	 proton structure function \protect\cite{WHIT,EMC}, 
	 smeared with a nucleon momentum distribution, $f(y)$, 
	 calculated in the
	 the on-mass-shell \protect\cite{FS78} and off-mass-shell 
	 \protect\cite{MST,MSTD} models.}    
\end{figure}

\begin{figure}
\centering{\ \psfig{figure=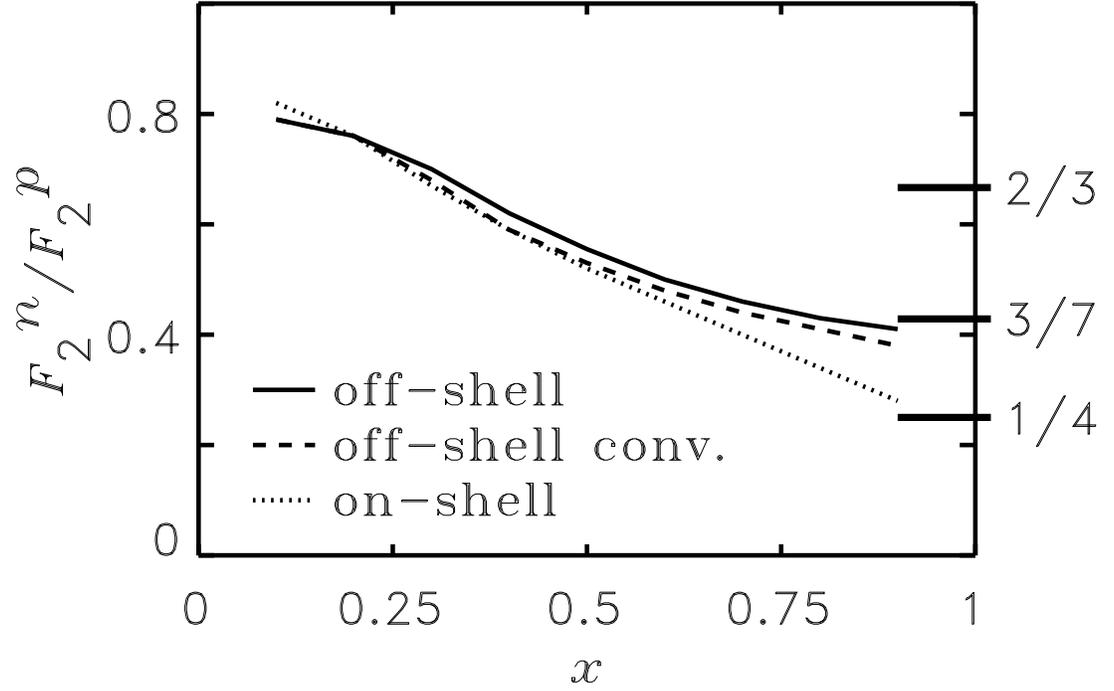,height=11cm}}
\caption{$F_2^n/F_2^p$ ratio as a function of $x$, for the 
         off-shell model (solid), off-shell model without 
         the convolution-breaking term (dashed), 
	 and the on-shell model (dotted).
	 On the right-hand axis are marked the $x \rightarrow 1$
       	 limits of the SU(6) symmetric model (2/3), and the predictions
         of the models of Refs.\protect\cite{CLO73,CAR75} (1/4) and 
	 \protect\cite{FJ,BBS} (3/7).}
\end{figure}

\begin{figure}
\centering{\ \psfig{figure=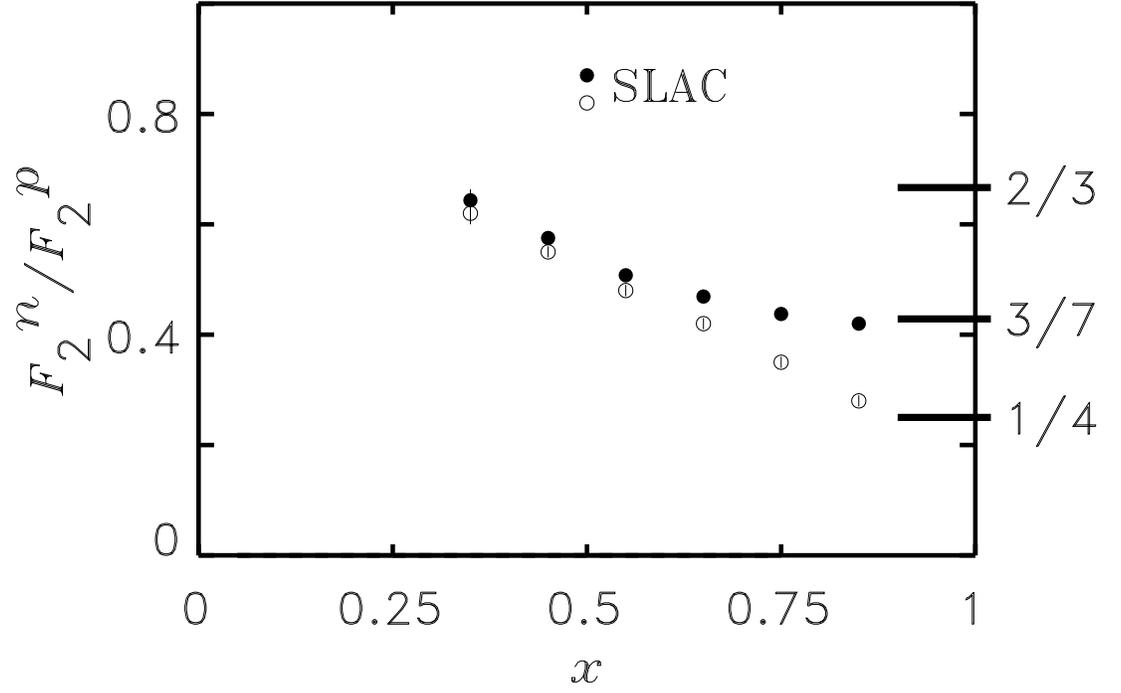,height=11cm}}
\caption{Deconvoluted $F_2^n/F_2^p$ ratio extracted from the
         SLAC $p$ and $D$ data \protect\cite{WHIT,GOMEZ}, 
	 at an average value of $Q^2 \approx 12$
	 GeV$^2$, assuming no off-shell effects (open circles), 
         and including off-shell effects (full circles).}
\end{figure}

\begin{figure}
\centering{\ \psfig{figure=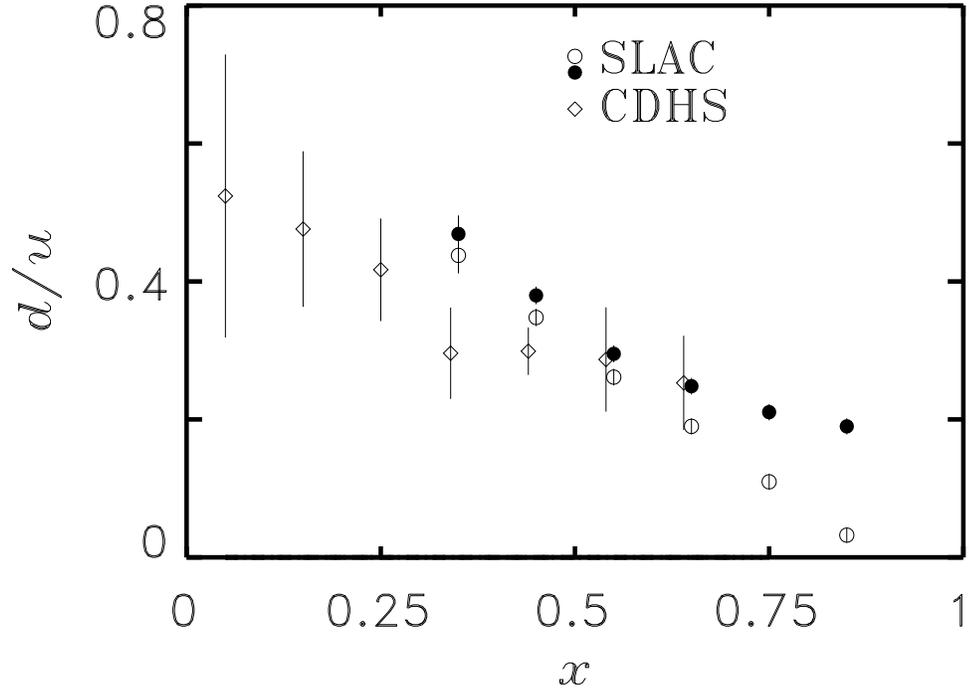,height=11cm}}
\caption{Extracted $d/u$ ratio, using the off-shell deuteron 
         calculation (full circles) and using on-shell kinematics
         (open circles).  Also shown for comparison is the ratio
         extracted from neutrino measurements by the CDHS collaboration
         \protect\cite{CDHS}.}    
\end{figure}

\end{document}